# Michael Ellis Fisher – CV and achievements

Amnon Aharony[1] and Mustansir Barma[2]

[1]School of Physics and Astronomy, Tel Aviv University,
Ramat Aviv, Tel Aviv 6997801, Israel
[2] TIFR Centre for Interdisciplinary Sciences, Tata Institute of Fundamental Research,
Hyderabad 500046, India

This text was supposed to be included in the book "*50 years of the renormalization group, Dedicated to the Memory of Michael E. Fisher*", edited by A. Aharony, O. Entin-Wohlman, D. Huse and L. Radzihovsky, World Scientific, Singapore (2024). It will be included in future printings and in the electronic version of the book.

# M. E. Fisher (1931-2021) -Curriculum vitae

Michael Ellis Fisher was born in Trinidad on 3 September 1931 and died in Maryland on 26 November 2021. He received his BSc from King's College London, where he also earned a PhD in physics in 1957. He was appointed to the faculty as a lecturer the following year, becoming a full professor in 1965. In 1966 he moved to Cornell University, where in 1973 he became the Horace White Professor of Chemistry, Mathematics and Physics, chairing the chemistry department from 1975 to 1978. Fisher was elected Secretary of the Cornell University Senate. In 1983, he was elected as a Foreign Associate Member of the National Academy of Sciences, Chemistry section. Since 1987 he was at the Institute for Physical Science and Technology at the University of Maryland, College Park, where he was a University System of Maryland Regents Professor, a Distinguished University Professor and Distinguished Scholar-Teacher until he retired in 2012.

Fisher was also a Fellow of the Royal Society, the American Academy of Arts and Sciences, the American Association for the Advancement of Science and the American Physical Society. Among his many awards were the Irving Langmuir Prize in Chemical Physics, the Boltzmann Medal in Thermodynamics and Statistical Mechanics and the Hildebrand Award of the American Chemical Society. Fisher received the inaugural American Physical Society Lars Onsager Prize for statistical physics in 1995. (His son Daniel received the same award in 2013.).

Michael Fisher's wife, Sorrel, died in 2016. Michael and Sorrel are survived by four children and seven grandchildren. Two of their sons are also theoretical physicists: Daniel S. Fisher is on the faculty at Stanford, while Matthew P. A. Fisher is a professor at the University of California, Santa Barbara.

**Wolf Prize**

In 1980, Fisher, together with Kenneth G. Wilson and Leo Kadanoff, won the Wolf Prize in Physics. Known both for his attention to detail and for his broad approach to understanding the world, Fisher was commended by the committee: "Professor Michael E. Fisher has been an extraordinarily productive scientist, and one still at the height of his powers and creativity. Fisher's major contributions have been in equilibrium statistical mechanics and have spanned the full range of that subject. He was mainly responsible for bringing together and teaching a common language to chemists and physicists working on diverse problems of phase transitions."

**Boltzmann Medal**

In 1983, Fisher was awarded the IUPAP Boltzmann Medal, with the following citation made by Cyril Domb:

"Michael Fisher first entered the field of statistical mechanics in the late 1950s. Within a few years, he had established a reputation as the leading authority in the field of critical phenomena, a position which he has maintained ever since. During the past two decades, he has been a major driving force behind the very great progress which has taken place. It is not possible in the short time available to do justice to the flood of papers with which Michael Fisher has been associated. Some of these have initiated new areas of research; for example, the exact susceptibility of the two-dimensional Ising model, correlation in the three-dimensional Ising model and critical scattering, renormalization of critical exponents resulting from hidden variables, finite-size scaling, the droplet model, partial differential approximants, the ANNNI model. Others, review articles, have become classics to which successive generations of graduate students and other researchers in the field have turned for guidance; for example, the Boulder lectures on critical phenomena, the 1964 Journal of Mathematical Physics review of correlation in fluids and magnets, the often-quoted 1967 review in Reports on Progress in Physics, and the 1973 Reviews of Modern Physics review of renormalization group. Each and every one of his papers contains new information of significance, and his collaborators will all verify that nothing is allowed to appear in print without Michael Fisher personally assuring himself that it measures up to his high standards. It would need quite an effort to list the research papers that have been sparked by footnotes in Michael Fisher's publications. Of not less importance than his publications has been the personal influence that he has exercised as a teacher on his many graduate students and collaborators; the invited lectures that he delivered so faultlessly and impeccably at countless national and international gatherings; the comments and criticisms which he has made during conference discussions (the atmosphere at conference is always more tense and exciting when Michael Fisher is present); and his many discussions with and directives to experimental workers in the field. From the historical point of view, the peak of his achievement to date has undoubtedly been the role that he played in the emergence of the renormalization group. Kenneth Wilson has stated publicly that all his knowledge of critical phenomena was acquired from Michael Fisher. Michael's presence at Cornell was an essential ingredient of his major achievement of the present era. The Boltzmann medal for 1983 is awarded to Michael Fisher for his many illuminating contributions to phase transitions and critical phenomena during the past 25 years."

**Lars Onsager Prize**

Fisher won the APS Lars Onsager Prize in 1995 "[f]or his numerous and seminal contributions to statistical mechanics, including but not restricted to the theory of phase transitions and critical phenomena, scaling laws, critical exponents, finite size effects, and the application of the renormalization group to many of the above problems".

# M. E. Fisher - Selected scientific achievements

The numbers in square brackets refer to Fisher's list of publications, which is included in the book and attached below. Many papers were written with collaborators. In those cases, `Fisher' implies `Fisher *et al.*'

**Exact and rigorous results**

Fisher obtained the partition function of close-packed hard **rigid dimers** on a square lattice exactly, by writing it as a Pfaffian [30, 31], and showed that the correlations of two monomers in a dimer lattice decay as a power law [41]. Exact closed expressions were derived for **decorated Ising models** [18], including a super-exchange model in which non-magnetic spins on bonds lead to antiferromagnetic couplings between spins [21, 24]. For the cluster size and percolation problem, Fisher found exact solutions on the Bethe lattice and obtained bounds for the critical probabilities on several lattices [28, 29]. Further, for classical or quantum systems, conditions on pair potentials were derived rigorously, which ensure **stability**, namely a lower bound on the ground state energy [56]. The existence of the **thermodynamic limit** was proved for the canonical free energy and the grand canonical pressure was calculated under general conditions on the attractive and repulsive parts of the interparticle potential [47]. Further, the existence and uniqueness of the limiting free energy per unit area for a planar wall was established under appropriate conditions on the potentials [166].

**Critical phenomena, scaling and critical exponents before the renormalization group**

The singularities in the behavior of statistical systems near the critical point were quantified by **defining critical exponents and determining their values** using various methods, including the analysis of high-temperature and low-temperature series expansions. Fisher was the first to show the need for the **correlation function exponent η** and to propose its relations with other critical exponents. The dimensional dependence of critical exponents was investigated by series expansion methods in [42]. His treatment of the **magnetic susceptibility** clarified the nature of singularities in ferromagnets and antiferromagnets. A detailed **analysis of experimental data** in nickel showed deviations from the mean field predictions and conformity with theoretical work on the Heisenberg model [50]. The **resistive anomaly** observed in metallic magnets was shown to arise from short-range fluctuations, thereby related to the specific heat anomaly [71]. Fisher introduced the **droplet model** of condensation, which showed the existence of an essential singularity at the phase transition point and led to **relations between critical exponents** [57]. The implications of a thermodynamic constraint such as a fixed concentration of impurities were shown to lead to the **Fisher renormalization** of critical exponents: their values depart in a specified way from their ideal values [72, 91]. His **review on critical phenomena** [66] addressed the behavior of a variety of systems, emphasizing the analogies between fluids, magnets, superfluids, and binary alloys, and included results for the critical behavior of Ising and *n*-vector models in different dimensions, rigorous inequalities for critical exponents, and the scaling hypothesis, which implied relations between exponents. In this review and in conferences [51, 58], lectures [57], and schools [3] in this period, Fisher's synthesis of extant knowledge advanced the field by bringing important unanswered questions into focus.

**Renormalization group and universality classes**

As already mentioned in the preface, the Wilson-Fisher paper [105] started the fifty years modern era of studying critical phenomena (and later many other systems, in physics and elsewhere) using the powerful **ε-expansion** for analytical solutions of Wilson's momentum-space **renormalization group**. This expansion used a completely original and unconventional idea in which the dimension *d*, or rather $\varepsilon = 4 - d$, the deviation from four dimensions, becomes an expansion parameter. That paper treated one- and two-component order-parameters, related to the universality classes of the Ising and the XY models with short-range interactions. This paper is mentioned in many papers in this book, in connection with recent developments. Fisher then generalized this analysis to many other universality classes, e.g., of isotropic **long-range interactions** [115], of **dipole-dipole interactions** [121, 123], and of the *n*-component order parameter with its **crossover to lower symmetries** [109, 134] (in parallel to Wegner). This work started much work on multicritical phase diagrams (see below). Fisher also started the discussions of the fixed point associated with **first order transitions** [215]. In a very important paper, Fisher discovered the **field theory for the Yang-Lee edge singularity**, with the associated expansion in $\varepsilon = 6-d$ dimensions [174]. This topic is still very active in connection to its relation to the branched polymers in *d-2* dimensions, which involves supersymmetry. It is notable that the ideas developed by MEF have become the core of modern theoretical physics, e.g., conformal field theory, applied in a large range of physical systems. Fisher continued working on the Yang-Lee singularity in many other contexts [171, 178, 179, 195, 321].

**Renormalization group and multicritical points**

Fisher wrote many papers that use the renormalization group and ε-expansions for studies of **competing order parameters**, the resulting multicritical points, phase diagrams, and the **crossover** from these points to other universality classes. Examples include the **tricritical point** [137, 145], to which Fisher also returned in [170, 173, 175, 178, 181] and in [198, 199] (where he used **real space renormalization group**), and **the bi- and the tetra-critical points** [138, 139, 147, 158, 163]. (The theoretical study of the bicritical point started with Fisher's mean-field study of super solids [116]). In particular, Fisher emphasized the role of **linear and non-linear scaling fields** in drawing bicritical phase diagrams [151], and in identifying **corrections to scaling** – using both renormalization group [193, 217] and other methods [211, 240, 242, 285], paving the way for accurate experimental fits. In [162], Fisher showed how the bicritical point can turn into a **triple point** at which three first-order lines meet. In [187], Fisher confirmed the universality of Heisenberg bicritical points. The slow **crossover** from mean-field theory to criticality was discussed in [260].

**Scaling and renormalization group for scattering**

Fisher introduced scaling into the analysis of **scattering functions**, related to the Fourier transforms of the **two-spin correlation functions**. This started with the original construction of approximate scaling functions, which obey the correct behavior for both short and long ranges [64, 102, 104]. These papers introduce the **new exponent η**, which was needed for the consistency of the scaling relations. Fisher then continued with deriving these scaling functions exactly, using the RG [132, 136]. Two-point correlation functions were also studied in [98, 99, 131].

**Critical behavior from series expansions**

Fisher's use of extrapolation methods to analyze **series expansions** yielded important results on the singular behavior near critical points. Besides early hints of **universality** through the independence of critical exponents on the type of lattice, this provided the backdrop for the first heuristic statement of a relation between critical exponents ($α'+2β+γ'=2$) [40]. Results were obtained for **self-avoiding walks** [19], **Ising** ferromagnets and antiferromagnets [17], **hard squares** [54], ***n*-vector models** [102, 109, 142], and **quenched random systems** [266, 269]. These methods also allowed critical behavior to be investigated in higher-dimensional lattice systems [42, 142], and yielded definitive results for **crossover induced by anisotropy** [134]. Besides, Fisher pioneered new methods of analysis. In particular, **inhomogeneous differential approximants** allowed for contributions from the analytic background [177] and yielded accurate estimates of the specific heat exponent [202]. **Partial differential approximants** for series with more than one variable [160, 212, 213, 257] provided an unbiased method to characterize **crossover** due to relevant variables [204, 237] including randomness [266], as well as **confluent corrections** to scaling [211, 242, 285] coming from irrelevant variables.

**Walls, wetting, and critical endpoints**

Fisher elucidated a multitude of interesting phenomena induced by the presence of **surfaces**, including surface transitions, **wetting,** and the behavior near a **critical endpoint**. In a near-critical binary fluid mixture, an effective force between parallel walls was predicted in [176], giving rise to the "**critical Casimir effect**", a new source of interactions between colloidal particles. This was checked theoretically by exact calculations for the two-dimensional Ising model in strip geometry [183, 185]. The shift of critical temperature in **thin films** was found in [222] as a function of film thickness and the properties of the walls. The density profile of an **adsorbed fluid** near a wall was matched to experiments in [282]. Wetting, prewetting, and surface transitions were considered in [221], and the phase diagram was shown to include a **wetting tricritical point**, while additional transitions were induced by **triple points** [232]. In [307, 310] it was shown that a fluctuation-induced first-order wetting transition can arise from a space-dependent stiffness. The exponents for the complete wetting transition in a random system were predicted in [251], and verified by numerical calculations in a **random Ising model** [277]. In the vicinity of a **critical endpoint,** the locus of first-order transitions was shown to develop a weak singularity whose form and amplitude ratio are determined by the specific heat along the critical locus [293, 295, 296]. The shape of the coexistence curve near critical endpoints was determined in [304], while **fluid interface tensions** were considered in [293, 294] and related experiments on He4 were analyzed in [308].

**Modulated commensurate phases in the ANNNI model**

Through systematic low-temperature expansions for the **axial next nearest neighbor Ising** (ANNNI) model with competing interactions, Fisher studied the vicinity of the multiphase point from which spring an infinite number of spatially **modulated commensurate phases [**186, 194, 196, 197, 200]. A similar approach was used to study **chiral Potts models** [205, 208]. The phase boundaries and corresponding domain wall energies were found explicitly at low temperatures, while crossover scaling properties were determined around the **decoupling points** of the ANNNI and four-state clock model [216].

**Quenched random systems**

Fisher was very interested in the **Harris criterion** for the relevance of **random bond** interactions at critical points. This crossover was studied (with a variety of methods) for dilute Bose fluids [239, 250], wetting [251, 277], Helium in Vycor [259], and Ising ferromagnets [266]. He also used *1/d* and high-temperature series to explore short-range Ising **spin glasses** [269].

**Finite-size scaling and numerical studies**

Fisher started working on **boundary and finite-size effects** in [69], with exact results for the free energies of boundaries and surfaces in two-dimensional lattices. The **scaling theory** for the **shift and rounding** of transitions in **finite thickness** systems was established and checked for the spherical model in [112, 113, 117]. The existence of such free energies, under specific conditions, was proven mathematically in [166, 172]. With Selke, Fisher performed **Monte Carlo simulations** for the spatially modulated phase in an Ising model [180] and for competing interactions in the two-dimensional Ising model [190]. Together with Privman, he established a full **finite-size scaling theory** for a variety of sample shapes, including the identification of **universal critical amplitudes**, and tested these predictions using renormalization group calculations, as well as both transfer matrix and Monte Carlo simulations [228, 235]. They also found the finite size scaling theory for **first-order rounding** [229, 241, 244]. In molecular dynamics studies of the critical properties of binary fluids, it was found essential to include finite-size effects to determine the dynamic critical behavior [392].

**Coulombic criticality and ionic fluids**

In the early 1990s, stimulated by experimental observations, several fundamental questions were raised on thermodynamics and critical behavior in systems that exhibit **long-range interactions**. Michael Fisher investigated the nature of criticality in **electrolytes** [312, 313, 315, 316]. The original ideas of Debye for electrolytes were significantly extended by accounting for the association processes, which allowed for the analysis of phase transitions and thermodynamics in general dimensions [317, 318, 319, 339, 340]. By developing a proper free energy description of the processes with long-range interactions, Fisher used complex calculations and advanced computer simulations to show that Ising criticality still applies to electrolytes, although the critical range is shorter than for typical short-range systems [324, 325, 329]. Significant progress was achieved in computer simulations of related processes [364, 367, 372]. Fisher also developed a completely new approach to understanding thermodynamics and criticality in lattice electrolyte systems [368, 377].

**Biophysics**

In early work related to biology, Fisher showed that the effect of self-avoidance on the denaturation transition in a double stranded polymer is to make it sharper, thus closer to experimental observation [61]. Later, Fisher modeled two-dimensional vesicles as closed self-avoiding walks and showed how the interplay between pressure and bending rigidity impacts their shapes, ranging from a circle to branched-polymer-like, as the pressure difference is varied [270, 287, 289, 298].

As a totally new direction for his research, at the end of 1990s Michael Fisher started to work on problems related to the understanding of **complex biological processes**. Stimulated by experimental single-molecule measurements that allowed monitoring of non-equilibrium biological molecules with unprecedented temporal and spatial resolutions, Fisher investigated the microscopic mechanisms of **motor proteins**, molecules that transform chemical energy into mechanical work [347, 351, 355, 393]. The idea was to utilize chemistry-inspired discrete-state stochastic models that should account for major properties of these biological systems. Fisher successfully developed a stochastic framework that allowed the convenient analysis of experimental observations [360, 371, 381, 389]. The important aspect of these studies was the ability to explain how external forces influence the **dynamics of biological molecular motors** [394, 396, 398, 409].

**Review papers**

Fisher wrote many highly cited **reviews**, which have been widely used to generate more research in the field. These included original **scaling theories** and identifications of **critical exponents** before the renormalization group [58, 66] and detailed explanations of the renormalization group [143, 333, 418].

**Contributions to other subjects**

The above list is surely not complete. Fisher also worked on **walks, self-avoiding-walks, and polymers** [11, 14, 27, 238, 306], including early work on the limit of **zero-order parameter components** [152], the critical behavior of **antiferromagnets**[17, 20, 34, 35, 36, 37], **quantum systems** [44, 62, 169, 271, 280, 290], **real space renormalization group** for three component models [198, 199], and much more.

**Acknowledgements**

We thank Anatoly Kolomeisky for writing items related to biology and ionic fluids, and Leo Radzihovsky for useful suggestions. MB acknowledges the support of the Indian National Science Academy and the Department of Atomic Energy, Government of India, under Project Identification No. RTI4007.

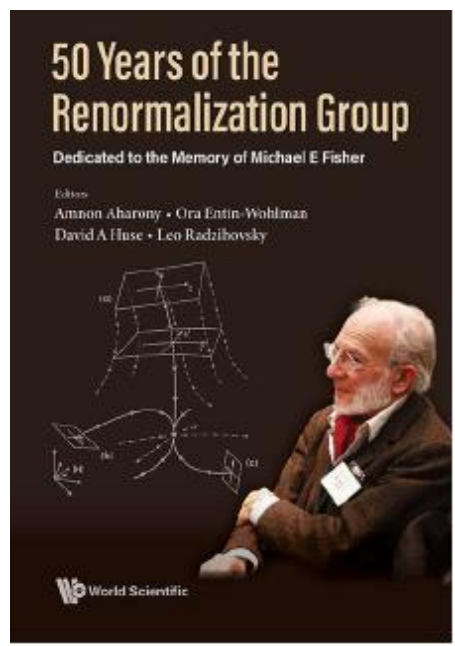

https://www.worldscientific.com/worldscibooks/10.1142/13571#t=aboutBook

# Michael E. Fisher – publications until 2017

This list was edited by Fisher himself and by his assistant. We thank Anne Suplee from UMD for sending it to us. We include here only the list of papers. Other publications are listed in the book.

1. Slide rule versus bead frame, J. Royal Air Force Technical College, 3, 53-60 (Oct. 1954).

2. The matrix approach to filters and transmission lines, Part I, Electronic Engineering 27, 198-204 (1955).

3. The matrix approach to filters and transmission lines, Part II, Electronic Engineering 27, 258-263 (1955).

4. Higher order differences in the analogue solution of partial differential equations, Journées Inter- nationales de Calcul Analogique - Sept. 55 - Proc. Internat. Conf. Analog. Computation (Brussels, Sept. 1955) pp. 208-213.

5. Higher order differences in the analogue solution of partial differential equations, J. Assoc. Comp. Mach. 3, 325-347 (1956).

6. On iterative processes and functional equations (C. Domb and M.E.F.) Proc. Camb. Phil. Soc. 52, 652-662 (1956).

7. On the continuous solution of integral equations by an electronic analogue, Proc. Camb. Phil. Soc. 53, 162-174 (1957).

8. Optimum design of quarter-squares multipliers with segmented characteristics, J. Sci. Insts. 34, 312-316 (1957).

9. Avoiding the need for dividing units in setting up differential analyzers, J. Sci. Insts. 34, 334-335 (1957).

10. A wide band analogue multiplier using crystal diodes and its application to the study of a non-linear differential equation, Electronic Engineering 29, 580-585 (1957).

11. On random walks with restricted reversals (C. Domb and M.E.F.) Proc. Camb. Phil. Soc. 54, 48-59 (1958).

12. Proposed methods for the analog solution of Fredholm's integral equation, J. Assoc. Comp. Mach. 5, 357-369 (1958).

(1990).

287. Tunable fractal shapes in self-avoiding polygons and planar vesicles (C.J. Camacho and M.E.F.) Phys. Rev. Lett. 65, 9-12 (1990).

288. Finite-size effects in fluid interfaces (M.P. Gelfand and M.E.F.) Physica A 166, 1-74 (1990).

289. Size of an inflated vesicle in two dimensions (A.C. Maggs, S. Leibler, M.E.F. and C.J. Camacho) Phys. Rev. A 42, 691-695 (1990).

290. Low-dimensional quantum antiferromagnets: Criticality and series expansions at zero temperature [extended abstract] Proc. Conf. "Frontiers of Condensed Matter Physics," Physica A 168, 22 (1990).

291. Interfaces, membranes: rough, smooth and interacting [summary] Proc. Conf. "Nonlinear Science: The Next Decade," Physica D 51, 498-500 (1991).

292. Universality and interfaces at critical end points (M.E.F. and P.J. Upton) Phys. Rev. Lett. 65, 2402-2405 (1990).

293. Critical endpoints, interfaces, and walls, Proc. Amer. Chem. Soc. "Symposium on Surfaces and Interfaces," Physica A 172, 77-86 (1991).

294. Fluid interface tensions near critical end points (M.E.F. and P.J. Upton) Phys. Rev. Lett. 65, 3405-8 (1990).

295. Phase boundaries near critical end points I. Thermo-dynamics and universality (M.E.F. and M.C. Barbosa) Phys. Rev. B 43, 11177-184 (1991).

296. Phase boundaries near critical end points II. General spherical models (M.C. Barbosa and M.E.F.) Phys. Rev. B 43, 10635-646 (1991).

297. Semiflexible planar polymeric loops (C.J. Camacho, M.E.F. and R.R.P. Singh) J. Chem. Phys. 94, 5693-5700 (1991).

298. Two-dimensional lattice vesicles and polygons (M.E.F., A.J. Guttmann and S.G. Whittington) J. Phys. A: Math. Gen. 24, 3095-3106 (1991).

299. Effective potentials, constraints, and critical wetting theory (M.E.F. and A.J. Jin) Phys. Rev. B 44, 1430-33 (1991).

300. Simulations of planar vesicles and their transitions (C.J. Camacho and M.E.F.) in "Computer Simulation Studies in Condensed Matter Physics IV," Eds. D.P. Landau, K.K. Mon and H.B. Schuttler (Springer Verlag, Berlin, 1993) pp. 189-193.

301. Effective interface Hamiltonians for short-range critical wetting (A.J. Jin and M.E.F.) Phys. Rev. B 47, 7365-7388 (1993).

302. Microcanonical density functionals for critical systems: An exact one-dimensional

317.   Coulombic criticality in general dimensions (Y. Levin, X.-J. Li and M.E.F.) Phys. Rev. Lett. 73, 2716-19 (1994); 75 [E] 3374 (1995).

318.   On the absence of intermediate phases in the two-dimensional Coulomb gas (M.E.F., X.-J. Li and Y. Levin) J. Stat. Phys. 79, 1-11; 81 [E] 865 (1995).

319.   Criticality in the hard-sphere ionic fluid (Y. Levin and M.E.F.) Physica A 225, 164-220 (1996).

320.   On the critical polynomial of the simple cubic Ising model, J. Phys. A: Math. Gen. 28, 6323-33 (1995).

321.   The universal repulsive-core singularity and Yang-Lee edge criticality (S.-N. Lai and M.E.F.) J. Chem. Phys. 103, 8144-55 (1995).

322.   Universal surface-tension and critical isotherm amplitude ratios in three dimensions (S.-Y. Zinn and M.E.F.) Physica A 226, 168-180 (1996).

323.   Foreword: About the author and the subject, to "The Critical Point" by C. Domb (Taylor and Francis, London, 1996), pp. xiii-xviii.

324(a) Density fluctuations in an electrolyte from generalized Debye-Hückel theory (B.P. Lee and M.E.F.) Phys. Rev. Lett. 76, 2906-10 (1996).

   (b) Inhomogeneous Debye-Hückel theory and density fluctuations near Coulombic criticality [abstract] (B.P. Lee and M.E.F.) Bull. Amer. Phys. Soc. 41 (1) 377 1996) I20 1.

325(a) Ginzburg criterion for Coulombic criticality (M.E.F. and B.P. Lee) Phys. Rev. Lett. 77, 3561-64 (1996).

   (b) The Ginzburg criterion for ionic criticality according to Debye-Hückel based theories [abstract] (M.E.F. and B.P. Lee) Bull. Amer. Phys. Soc. 41 (1) 377 (1996) I20 2.

326(a) Dipolar-ion pairs and Debye-Hückel theory: The charging process and reciprocity [abstract](B.P. Lee and M.E.F.) Bull. Amer. Phys. Soc. 41 (1) 356 (1996) I7 11.
   (b) Generalized Debye-Hückel theory and ionic criticality (B.P. Lee and M.E.F.) [in preparation].

327.   Criticality in Gaussian-molecule mixtures (S.-N. Lai and M.E.F.) Molec. Phys. 88, 1373-1397 (1996).

328.   Renormalized coupling constants and related amplitude ratios for Ising systems (S.-Y. Zinn, S.-N. Lai, and M.E.F.) Phys. Rev. E 54, 1176-1182 (1996).

329(a) Coulombic criticality in general dimensions: Progress and challenges [extended abstract] Proc. Hayashibara Forum '95: "Coherent Approaches to Fluctuations," Eds. M. Suzuki and N. Kawashima (World Scientific Publ. Co., Singapore, 1996) pp. 2-5.

329(b)   The nature of criticality in ionic fluids, Proc. Third Liquid Matter Conference, J. Phys.: Condens. Matter $\underline{8}$, 9103-9109 (1996).

   (c)   Criticality in hard-core ionic systems: Progress and challenges [extended abstract] Proc. Inauguration Conf. Asia Pacific Center for Theoretical Physics, "Current Topics in Physics," Eds. Y.M. Cho, J.B. Hong and C.N. Yang (World Scientific Publ. Co., Singapore, 1998) pp. 100-105.

330.   The superfluid transition in a dilute Bose gas: Experiments and theory [abstract] Proc. Inauguration Conf. Asia Pacific Center for Theoretical Physics, "Current Topics in Physics," Eds. Y.M. Cho, J.B. Hong and C.N. Yang (World Scientific Publ. Co., Singapore, 1998) p. 175.

331.   Right and wrong near critical endpoints (M.E.F. and Y.C. Kim) J. Chem. Phys. $\underline{117}$, 779-787 (2002).

332(a)   Prewetting transitions in a near-critical metallic vapor (V.F. Kozhevnikov, D.I. Arnold, S.P. Naurzakov and M.E.F.) Phys. Rev. Lett. $\underline{78}$, 1735-1738 (1997).
   (b)   Prewetting phenomena in mercury vapor (V.F. Kozhevnikov, D.I. Arnold, S.P. Naurzakov, and M.E.F.) Fluid Phase Equilibria, $\underline{150\text{-}151}$, 625-632 (1998).

333.   Renormalization group theory: Its basis and formulation in statistical physics, Rev. Mod. Phys. $\underline{70}$, 653-681 (1998).
REPRINTED (with revisions) in "Conceptual Foundations of Quantum Field Theory," Ed., T.Y. Cao (Cambridge Univ. Press, 1998) Part IV, Chap. 8, pp. 89-135.

334(a)   Critique of primitive model electrolyte theories (D.M. Zuckerman, M.E.F and B.P. Lee) Phys. Rev. E $\underline{56}$, 6569-6580 (1997).
   (b)   Critique of electrolyte theories using thermodynamic bounds (M.E.F., D.M. Zuckerman and B.P. Lee) in "Strongly Coupled Coulomb Systems," Proc. Conf. held in Boston College, August 1997, Eds. G.J. Kalman, J.M. Rommel and K.B. Blagoev (Plenum Press, New York 1998) pp. 415-418.

335(a)   Charge oscillations in Debye-Hückel theory (B.P. Lee and M.E.F.) Europhys. Lett. $\underline{39}$, 611-616 (1997).
   (b)   Charge oscillations in electrolytes from generalized Debye-Hückel theory [abstract] (B.P. Lee and M.E.F.) Bull. Amer. Phys. Soc. $\underline{42}$ (1) 725 (1997) Q16 9.

336(a)   Diverging correlation lengths in electrolytes: Exact results at low densities (S. Bekiranov and M.E.F) Phys. Rev. E $\underline{59}$, 492-511 (1999).
   (b)   Exact density correlation length for electrolytes at low densities [abstract] (S. Bekiranov and M.E.F.) Bull. Amer. Phys. Soc. $\underline{42}$ (1) 725 (1997) Q16 8.

337.   Electrolyte criticality and generalized Debye-Hückel theory (M.E.F., B.P. Lee and S. Bekiranov) in "Strongly Coupled Coulomb Systems," Proc. Conf. held in Boston College, August 1997, Eds. G.J. Kalman, J.M. Rommel and K. B. Blagoev (Plenum Publ. Corp., 1998) pp. 33-41.

338.   Exact thermodynamic formulation of chemical association (M.E.F. and D.M.

4558-61 (2000).

362. Asymmetric primitive-model electrolytes: Debye-Hückel theory, criticality and energy bounds (D.M. Zuckerman, M.E.F. and S. Bekiranov) Phys. Rev. E. 64, 011206:1-13 (2001).

363(a) The heat capacity of the restricted primitive model electrolyte (E. Luijten, M.E.F. and A.Z. Panagiotopoulos) J. Chem. Phys. 114, 5468-71 (2001).
 (b) Criticality and charge fluctuations in the restricted primitive model electrolyte [abstract] (E. Luijten, M.E.F. and A.Z. Panagiotopoulos) Bull. Amer. Phys. Soc. 46 (1) 71 (2001) A11 4.

364. Precise simulation of criticality in asymmetric fluids (G. Orkoulas, M.E.F. and A.Z. Panagiotopoulos) Phys. Rev. E 63, 051507:1-14 (2001).

365. The critical locus of a simple fluid with added salt (Y.C. Kim and M.E.F.) J. Phys. Chem. B 105, 11785-95 (2001).

366. Phase transitions in 2:1 and 3:1 hard-core model electrolytes (A.Z. Panagiotopoulos and M.E.F.) Phys. Rev. Lett. 88, 045701:1-4 (2002).

367(a) Universality class of criticality in the restricted primitive model electrolyte (E. Luijten, M.E.F. and A. Z. Panagiotopoulos) Phys. Rev. Lett. 88, 185701:1-4 (2002).

367(b) How to simulate fluid criticality: The simplest ionic model has Ising behavior but the proof is not so obvious! Proc. Internat. Conf. Theoretical Physics: TH-2002, UNESCO, Paris, 22-27 July 2002, Ann. Henri Poincaré 4, 413-416 (2002).

367(c) Charge fluctuations and criticality in the restricted primitive model electrolyte [abstract] (E. Luijten, M.E.F. and A.Z. Panagiotopoulos) in "STATPHYS 21 Conference Abstracts," Eds, D. López, M. Barbosa and A. Robledo (IUPAP, Cancun, Mexico, 2001) p. 66.

368. Lattice models of ionic systems (V. Kobolev, A.B. Kolomeisky and M.E.F.) J. Chem. Phys. 116, 7589-7598 (2002).

369. Criticality in charge asymmetric ionic fluids (J.-N. Aqua, S. Banerjee and M.E.F.) [arXiv:cond-mat/0410692] Phys. Rev. E 72, 041501:1-25(2005).

370. Motion of kinesin on a microtubule [abstract] (M.E.F and A.B. Kolomeisky) Biophys. J. 82, 62a (2002) 305-Pos.

371(a) A simple kinetic model describes the processivity myosin-V (A.B. Kolomeisky and M.E.F.) Biophys. J. 84, 1642-1650 (2003).
 (b) A simple stochastic model can explain the motility of myosin V molecules [abstract] (A.B. Kolomeisky and M.E.F.) Biophys. J. 82, 15a (2002) 71-Plat.

372. Screening in ionic systems: Simulations for the Lebowitz length (Y.C. Kim, E. Luijten and M.E.F.) Phys. Rev. Lett. 95, 145701:1-4 (2005).

373. Asymmetric fluid criticality I. Scaling with pressure mixing (Y.C. Kim, M.E.F. and G. Orkoulas) Phys. Rev. E 67, 061506:1-21 (2003).

384^    Yang-Yang anomalies and coexistence curve diameters: Simulation of asymmetric fluids (Y.C. Kim) [arXiv:cond-mat/0503480] Phys. Rev. E $\underline{71}$, 051501:1-16 (2005).

385.    Interfaces: Fluctuations, interactions and related transitions, Chap. 2 in "Statistical Mechanics of Membranes and Surfaces" (Second Edition, enlarged and revised) Eds. D. Nelson, T. Piran and S. Weinberg (World Scientific Publ., Singapore, 2004) pp. 19-47: see [281].

386.    How multivalency controls ionic criticality (M.E.F., J.N. Aqua and S. Banerjee) [arXiv:cond-mat/0507077] Phys. Rev. Lett. $\underline{95}$, 135701:1-4 (2005).

387.    Interfacial tensions near critical endpoints: Experimental checks of EdGF theory (S.-Y. Zinn and M.E.F.) [arXiv:cond-mat/05043857] Molec. Phys. $\underline{103}$, 2927-2942 (2005): special issue in honor of B. Widom.

388.    Singular coexistence-curve diameters: Experiments and simulations (Y.C. Kim and M.E.F.) [arXiv:cond-mat/ 0507369] Chem. Phys. Lett. $\underline{414}$, 185-192 (2005).

389(a)  Kinesin crouches to sprint but resists pushing (M.E.F. and Y.C. Kim) Proc. Natl. Acad. Sci. USA 102, 16209-14 (2005).

  (b)   Kinesin binds ATP, crouches and swings 8 nm; But rebuffs help [abstract] (Y.C. Kim and M.E.F.) Biophys. J. $\underline{88}$, 649a (2005) 3189-Pos.

  (c)   Kinesin crouches before sprinting and resists forward and leftward loading [abstract] (Y.C. Kim and M.E.F.) Proc. 228[th] ACS National Meeting, "Biophysical Chemistry and Novel Imaging of Single Molecules and Single Cells," Division of Physical Chemistry, Philadelphia, PA (25 August 2004.)

390.    Universality of ionic criticality: Size- and charge-asymmetric electrolytes (Y.C. Kim, M.E.F. and A.Z. Panagiotopoulos) Phys. Rev. Lett. $\underline{95}$, 195703:1-4 (2005).

391(a)  Static and dynamic critical behavior of a symmetrical binary fluid: A computer simulation (S.K. Das, J. Horbach, K. Binder, M.E.F. and J.V. Sengers) [arXiv: cond-mat/0603587 22 Mar 2006] J. Chem. Phys. $\underline{125}$, 024506:1-11 (2006).

  (b)   Interdiffusion in Critical Binary Mixtures by Molecular Dynamics Simulation (K. Binder, S.K. Das, M.E.F., J. Horbach, and J.V. Sengers) in "Diffusion Fundamentals II; L'Aquila 2007," Eds. S. Brandani, C. Chmelik, J. Kärger and R. Volpe (Leipziger Universitätsverlag, Leipzig, 2007) pp. 120-131.

392.    Critical dynamics in a binary fluid: Simulations and finite-size scaling (S.K. Das, M.E.F., J.V. Sengers, J. Horbach and K. Binder) Phys. Rev. Lett. $\underline{97}$, 025702:1-4 (2006).

393.    Molecular Motors: A theorist's perspective (A.B. Kolomeisky and M.E.F.) Ann. Rev. Phys. Chem. $\underline{58}$, 675-695 (2007).

394(a)  Backstepping, hidden substeps, and conditional dwell times in molecular motors (D. Tsygankov, M. Lindén and M.E.F.) [arXiv:q-bio BM/0611051] Phys. Rev. E $\underline{75}$, 021909:1-16 (2007).

(b)     Back-stepping, dwell times, and hidden substeps in molecular motors [abstract] (D. Tsygankov, M. Lindén and M.E.F.) Biophys. J. 90 (Jan. 2007) 496a: 2369-Pos.

395(a)  Charge fluctuations and correlation lengths in finite electrolytes (Y.C. Kim and M.E.F.) Phys. Rev. E 77, 051502:1-7 (2008).

(b)     Charge Fluctuations and correlations in Finite electrolytes, contributed talk (Y.C. Kim and M.E.F.) American Physical Society Meeting, Baltimore, about 16 March 2006.

396(a)  Mechanoenzymes under superstall and large assisting loads reveal structural features (D. Tsygankov and M.E.F.) Proc. Natl. Acad. Sci. USA 104, 19321-326 (2007).

(b)     Superstall and assisting-load velocities of motor proteins can reveal mechano-chemical structure [abstract] (D. Tsygankov and M.E.F.) Biophys. J. 90 (Jan. 2007) 497a: 2371-Pos.

397.    Simulating critical dynamics in liquid mixtures: Short-range and long-range contributions (S.K. Das, J.V. Sengers and M.E.F) J. Chem. Phys. 127, 144506:1-5 (2007).

398.    Kinetic models for mechanoenzymes: Structural aspects under large loads (D. Tsygankov and M.E.F) J. Chem. Phys. 128, 015102:1-12 (2008).

399.    Boltzmann Award Session; Remarks on the Boltzmann Medalists (*Kurt Binder* and *Giovanni Gallavotti*) Proc. STATPHYS 23, Euro. Phys. J. B 64, 301, 303-5 (2008).

400(a)  Velocity statistics distinguish quantum turbulence from classical turbulence (M.S. Paoletti, M.E.F., K.R. Sreenivasan and D.P. Lathrop) Phys. Rev. Lett. 101, 154501:1-4 (2008); designated an Editor's Selection and selected for a Viewpoint in *Physics Today*.

(b)     Quantum turbulence [abstract] (D.P. Lathrop, M.S. Paoletti, M.E.F., K.R. Sreenivasan) Bull. Amer. Phys. Soc. 54 (19) ET 1, 132 (2009).

(c)     Velocity statistics in superfluid and classical turbulence [abstract](K.R. Sreenivasan, D.A. Donzis, M.E.F., D.P. Lathrop, M.S. Paoletti, and P.K. Young) Bull. Amer. Phys. Soc. 54 (19) ET 2, 132 (2009).

401(a)  Comment on a recent conjectured solution of the three-dimensional Ising Model (F.Y. Wu, B.M. McCoy, M.E.F. and L. Chayes) Phil. Mag. 88, 3093-95 (2008).

(b)     Rejoinder to the Response to 'Comment on a recent conjectured solution of the three-dimensional Ising Model' (F.Y. Wu, B.M. McCoy, M.E.F. and L. Chayes) Phil. Mag. 88, 3103 (2008).

(c)     Erratum for 'Comment' and 'Rejoinder,'(F.Y.Wu *et al.*) Phil. Mag. 89, 195 (2009).

402.    Criticality in multicomponent spherical models: Results and cautions (J.-N. Aqua and M.E.F) Phys. Rev. E 79, 011118:1-13(2009).

403.    Critical charge and density coupling in ionic spherical models (J.-N. Aqua and M.E.F.) [in preparation].

404.    Reconnection dynamics for quantized vortices (M.S. Paoletti, M.E.F. and D.P.

    

Chap. 10, pp. 167-198.

(b) 　　Statistical Physics in the Oeuvre of Chen Ning Yang [Invited review] Int. J. Modern
　　　　　　Phys. B **29**, 1530013:
　　　　　　1-31(2015).

(c) 　　Statistical Physics in the Oeuvre of Chen Ning Yang, Asia
　　　　Pacific Newsletter 5, 22-30 (2016).

422. Michael Ellis Fisher: pp. 2-74, Commentary, pp. 6-13 (with
　　　　　　Papers [45] and [384] and List of Publications in Wolf Prize in Physics, Ed. T.
　　　　　　Piran (World Scientific Publ. Co., Singapore, 20